\documentclass[aps, prx, twocolumn, superscriptaddress, ]{revtex4-1}
\usepackage{amsmath,amssymb,amsfonts,amsbsy}
\usepackage{graphicx}
\usepackage{subfig}
\graphicspath{{./figures/}}
\usepackage{dcolumn}
\usepackage{bm}
\usepackage{multirow}
\usepackage{mathtools}
\usepackage{array}
\usepackage{color}
\usepackage[normalem]{ulem}
\usepackage[per-mode=symbol]{siunitx}
\usepackage{upgreek}

\usepackage[labelfont=bf]{caption}
\usepackage[figurename=Fig.]{caption}

\DeclareSIUnit \belm {Bm}

\usepackage[colorlinks=true,breaklinks=true,allcolors=blue]{hyperref} 
\usepackage[capitalise,nameinlink]{cleveref} 
\usepackage{booktabs}
\usepackage[utf8]{inputenc}
\usepackage[T1]{fontenc}

\newcommand{\app}[1]{\hyperref[app:#1]{Appendix~\ref*{app:#1}}}

\def\@setaltaffiliation{\vspace{-\baselineskip}\def\altaffiliation##1{\@par##1\@addpunct.}\altaffiliationes}
\def\@setaltaffiliation{\vspace{-\baselineskip}\def\altaffiliation##1{\@par##1\@addpunct.}\altaffiliationes}

\RequirePackage{lineno}
\setpagewiselinenumbers

\begin{document}
\title{Memristor-based cryogenic programmable DC sources for scalable in-situ quantum-dot control}

\author{Pierre-Antoine~Mouny}
\altaffiliation{Correspondence author: \\pierre-antoine.mouny@usherbrooke.ca}
\author{Yann~Beilliard}
\affiliation{Institut Interdisciplinaire d’Innovation Technologique (3IT), Université de Sherbrooke, Sherbrooke, Québec J1K 0A5, Canada}
\affiliation{Laboratoire Nanotechnologies Nanosystèmes (LN2) – CNRS UMI-3463 – 3IT, Sherbrooke, Québec J1K 0A5, Canada}
\affiliation{Institut quantique (IQ), Université de Sherbrooke, Sherbrooke, Québec J1K 2R1, Canada
}
\author{Sébastien~Graveline}
\affiliation{Institut Interdisciplinaire d’Innovation Technologique (3IT), Université de Sherbrooke, Sherbrooke, Québec J1K 0A5, Canada}
\affiliation{Laboratoire Nanotechnologies Nanosystèmes (LN2) – CNRS UMI-3463 – 3IT, Sherbrooke, Québec J1K 0A5, Canada}
\author{Marc-Antoine~Roux}
\affiliation{Institut quantique (IQ), Université de Sherbrooke, Sherbrooke, Québec J1K 2R1, Canada
}
\author{Abdelouadoud~El~Mesoudy}
\author{Raphaël~Dawant}
\author{Pierre~Gliech}
\author{Serge~Ecoffey}
\affiliation{Institut Interdisciplinaire d’Innovation Technologique (3IT), Université de Sherbrooke, Sherbrooke, Québec J1K 0A5, Canada}
\affiliation{Laboratoire Nanotechnologies Nanosystèmes (LN2) – CNRS UMI-3463 – 3IT, Sherbrooke, Québec J1K 0A5, Canada}
\author{Fabien~Alibart}
\affiliation{Institut Interdisciplinaire d’Innovation Technologique (3IT), Université de Sherbrooke, Sherbrooke, Québec J1K 0A5, Canada}
\affiliation{Laboratoire Nanotechnologies Nanosystèmes (LN2) – CNRS UMI-3463 – 3IT, Sherbrooke, Québec J1K 0A5, Canada}
\affiliation{Institute of Electronics, Microelectronics and Nanotechnology (IEMN), Université de Lille, 59650 Villeneuve d’Ascq, France
}
\author{Michel~Pioro-Ladrière}
\affiliation{Laboratoire Nanotechnologies Nanosystèmes (LN2) – CNRS UMI-3463 – 3IT, Sherbrooke, Québec J1K 0A5, Canada}
\affiliation{Institut quantique (IQ), Université de Sherbrooke, Sherbrooke, Québec J1K 2R1, Canada
}
\author{Dominique~Drouin}
\affiliation{Institut Interdisciplinaire d’Innovation Technologique (3IT), Université de Sherbrooke, Sherbrooke, Québec J1K 0A5, Canada}
\affiliation{Laboratoire Nanotechnologies Nanosystèmes (LN2) – CNRS UMI-3463 – 3IT, Sherbrooke, Québec J1K 0A5, Canada}
\affiliation{Institut quantique (IQ), Université de Sherbrooke, Sherbrooke, Québec J1K 2R1, Canada
}

\date{\today}
\begin{abstract}

Current quantum systems that are based on spin qubits are controlled by classical electronics located outside the cryostat at room temperature. This approach creates a major wiring bottleneck, which is one of the main roadblocks toward truly scalable quantum computers. Thus, we propose a scalable memristor-based programmable DC source that can be used to perform biasing of quantum dots inside the cryostat (i.e., \textit{in-situ}). This novel cryogenic approach would enable us to control the applied voltage on the electrostatic gates by programming the resistance of the memristors, thus storing in the latter the appropriate conditions to form the quantum dots. In this study, we first demonstrate multilevel resistance programming of TiO$_{\textrm{2}}$-based memristors at \SI{4.2}{\kelvin}, which is an essential feature to achieve voltage tunability of the memristor-based DC source. We then report hardware-based simulations of the electrical performance of the proposed DC source. A cryogenic TiO$_{\textrm{2}}$-based memristor model fitted on our experimental data at \SI{4.2}{\kelvin} was used to show a \SI{1}{\volt} voltage range and \SI{100}{\micro\volt} in-situ memristor-based DC source. Finally, we simulate the biasing of double quantum dots, enabling sub-2 minutes in-situ charge stability diagrams. This demonstration is a first step towards more advanced cryogenic applications for resistive memories, such as cryogenic control electronics for quantum computers. 
\end{abstract}
\maketitle

\section{Introduction}

\begin{figure*}[htbp]
\centering
  \includegraphics[width=1\linewidth]{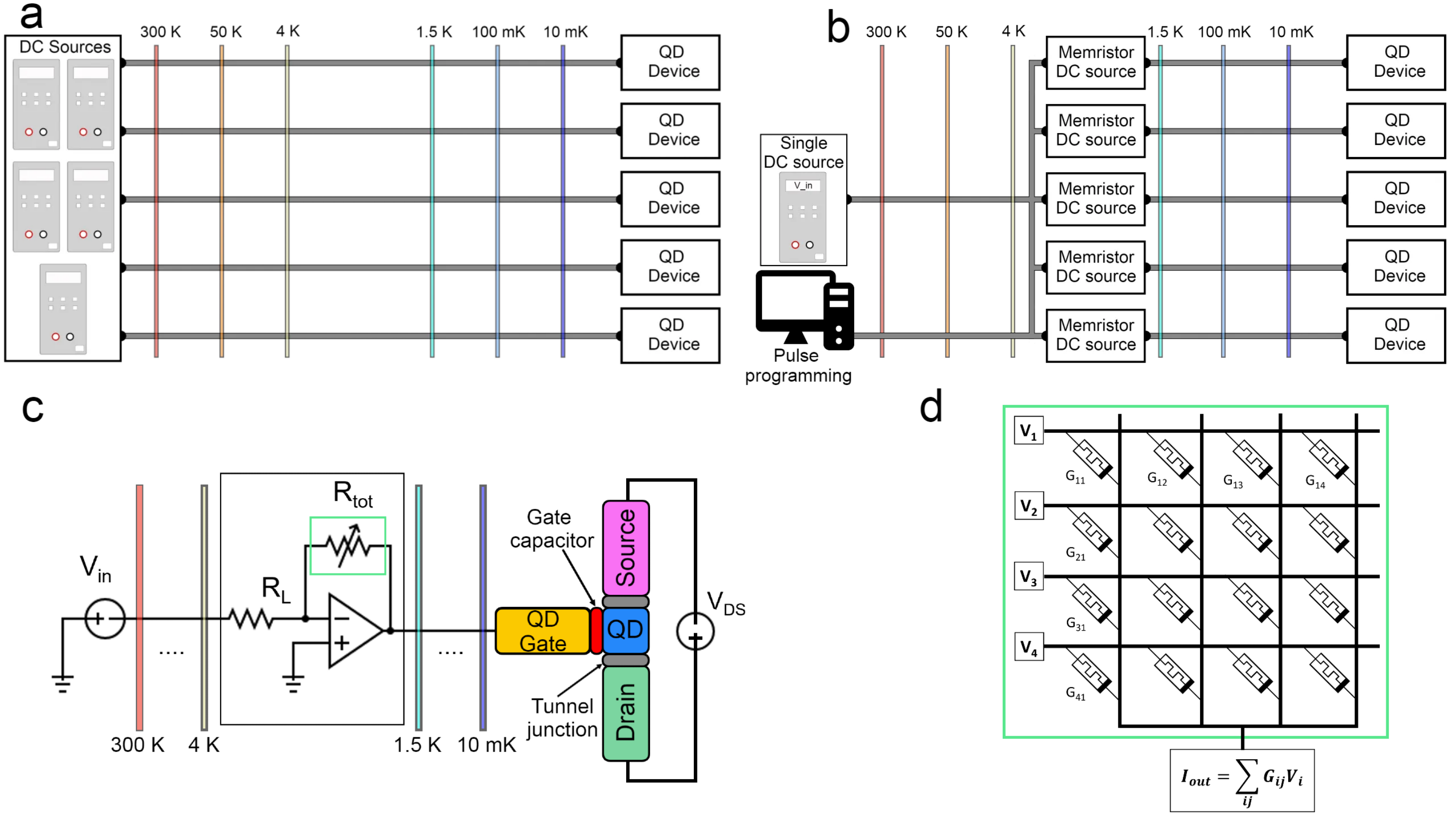}

\caption{\textbf{The memristor-based programmable DC source} \textbf{a.} Schematic view of a quantum dot based quantum computer that is strictly limited to the QD biasing. The qubit chip is placed at the lower stage of a dilution refrigerator at \SI{10}{\milli\kelvin}. The DC signals that are required to control the qubits are generated by DC sources at room temperature and routed by coaxial cables to base temperature. Attenuators and filters are fitted on the routing cables to improve the noise to signal ratio. Each quantum dot requires up to six DC cables, each of which is often connected to a dedicated tunable DC source at \SI{300}{\kelvin}, which leads to a wiring bottleneck. \textbf{b.} Schematic view of a quantum dot based quantum computer using the proposed memristor-based DC source at \SI{4}{\kelvin} and supplied by a single DC source delivering a fixed voltage $V_{\textrm{in}}$, replacing all of the the required DC sources at room temperature. An additional RF line is required to tune the memristors. \textbf{c.} Zoom-in on a single DC line supplying the voltage bias to a single QD gate. The green-framed variable resistor is the memristor circuit that is used as a tunable feedback resistor in a programmable gain amplifier circuit (framed in grey). \textbf{d.} Circuit diagram of a memristor-based variable resistor. Memristors are fabricated at the intersection of word lines and bit lines in a crossbar array. This architecture is able to address a larger number of memristors. By individually tuning the resistance of each memristor, we can vary the total resistance $R_{\mathrm{tot}}$ of the crossbar and thus the output voltage $V_{\textrm{out}}$ of the DC source.}
\label{fig:setup}
\end{figure*}

Silicon quantum dots (QDs) are among the leading candidates for large scale integration of qubits given their compatibility with industrial semiconductor fabrication processes \cite{Pillarisetty_2018qubits, zwerver2021qubits}, which may enable mass production and easier co-integration with control electronics. Moreover, semiconductor qubits benefit from a small qubit pitch, a long coherence time, and high gate fidelity \cite{Veldhorst2014, Yoneda2017, Watson2018, Zajac2017}. However, quantum error correction and millions of physical qubits will be required to enable key industrial applications, such as encryption, quantum machine learning, or drug synthesis. The current approach to realize a quantum dot-based computer requires the qubit chips to be operated at \SI{1}{\kelvin} or below \cite{Yang2020, Petit2020}, while its control charge carriers for state manipulation and readout are placed outside of the dilution refrigerator (see \cref{fig:setup}a). 
In particular, the number of quantum-dot gates that are required to properly confine the charge carriers can range from one to up to six for specific quantum dot designs \cite{Geck2019}, and as many dedicated DC sources and cables per qubit.
Although the control approach that is depicted in \cref{fig:setup}a has enabled important demonstrations over the past 20 years, it will become very cumbersome for few hundred qubits and may even be unachievable for the million of physical qubits that are required for fault tolerant quantum computing \cite{Bardin2018}. To avoid this wiring bottleneck, integrated cryogenic control platforms are being developed to control a larger number of spin qubits \cite{Vandersypen2017, Geck2019, Bardin2018,ruffino2021integrated, Xue2021, Pauka2021}, while limiting the control electronics required at room temperature (See \cref{fig:setup}b).
\linebreak

To identify and reach the correct electronic regime, it is necessary to individually tune all of the quantum-dot gate biases (see \cref{fig:setup}a and b) using DC signals, typically in the $\pm$\SI{1}{\volt} range with at least a \SI{100}{\micro\volt} resolution, while maintaining a thermal budget in the order of a milliwatt or below per DC source \cite{Vandersypen2017, Geck2019}. A first integrated solution based on switched-capacitor circuits and charge-locking, which is conceptually close to dynamic random access memories (DRAMs), has been proposed to perform simultaneous charge carrier confinement for several quantum dots\cite{XuSwitchedCapa2020, ruffino2021integrated}. However, this approach requires qubits that have similar electrical behaviour, which has yet to be demonstrated for silicon spin qubits. To maintain the desired bias voltage applied to the quantum-dot gate, the switched-capacitor charge needs to be refreshed periodically. Meanwhile, TiO$_{\mathrm{2}}$-based memristors \cite{Chua1971, Strukov2008} exhibit promising features, such as non-volatility, multilevel resistance state programming \cite{Alibart2012}  and CMOS-compatible fabrication processes \cite{mesoudy2021CMOS}. The memristor is used as the core component of the DC source. Its resistance value depends on its internal state. This resistance state has the particularity of being adjustable by external stimuli, such as DC sweeping  or voltage pulses \cite{LeCun2015, Amirsoleimani2020}, which enables the voltage programmability of the DC source. Moreover, the conduction mechanisms and temperature dependent DC behavior of memristors based on transition metal oxides have been studied at cryogenic temperatures down to \SI{1.5}{\kelvin} \cite{Beilliard2020, Fang2015, Blonkowski2015,Pickett2011, Zhang2014,Alagoz2019, Voronkovskii2019}, hinting at hybrid CMOS-memristor control circuits for quantum systems. The co-integration of a memristor-based biasing circuit and QDs will benefit from the recently demonstrated operation of spin qubits at \SI{1}{\kelvin} \cite{Yang2020, Petit2020} relaxing the power dissipation restrictions. This approach would make it possible to control the applied voltage on the gates by changing  the resistance of the memristors in a non-volatile manner, thus storing in the latter the appropriate conditions for the QDs to be formed. This on-chip co-integration would require a single I/O to bias several quantum dots, thus paving the way for scalable quantum computers on silicon.  

In this paper, we propose a memristor-based DC source set to operate at \SI{4}{\kelvin} to first control the voltage bias applied to the gates of a silicon quantum dot cooled to $\sim$\SI{10}{\milli\kelvin}. However, the perspective of spin qubits at \SI{4.2}{\kelvin} in the near future would allow a co-integration with the memristor-based DC source enabling scalability as both technologies are CMOS-compatible \cite{mesoudy2021CMOS, Maurand2016}. Using a feedback resistance tuning algorithm, we experimentally demonstrate the multilevel resistance programmability of CMOS-compatible Al$_{\mathrm{2}}$O$_{\mathrm{3}}$/TiO$_{\mathrm{2-x}}$ memristor crosspoints \cite{mesoudy2021CMOS} at \SI{4.2}{\kelvin}. In addition, we conduct extensive hardware-based simulations of a memristor-based programmable gain amplifier (PGA) to optimize the design of the memristor feedback resistor, and show that this approach meets the voltage range and resolution criteria for \textit{in-situ} biasing of quantum dots.

\section{Concept}

\Cref{fig:setup}a shows a typical experimental setup for solid-state spin qubits that operate at a temperature of \SI{1}{\kelvin} or below with optimal operation in the tens of millikelvins. DC biases are generated by rack-sized electronics at room temperature and transported by DC lines to the qubits in the dilution refrigerator. This approach imposes a limit on the number of biasing lines that can be integrated. We propose a programmable gain amplifier (PGA) as a biasing solution where the variable feedback resistor is based on a memristor circuit (See \cref{fig:setup}c, d). The resistance of each memristor can be finely tuned individually due to the analog behaviour of resistive memories. Write pulses are used to accurately program the resistance of the memristors. In the context of the PGA, these pulses can be applied using a low-dissipation cryogenic FPGA \cite{Artix_cryo_charbon}. A single DC source that provides a fixed voltage at room temperature is required to supply multiple PGA circuits placed at the \SI{4}{\kelvin}-stage (See \cref{fig:setup}b, c) limiting the control electronic that is required at room temperature and thus limiting the number of cables going in the cryostat.\linebreak

Based on the circuit parameters in \cref{fig:setup}c, we can fully determine the theoretical electrical characteristics of the memristor-based DC source. The output voltage of the PGA is determined by the amplification factor $A_R=R_{\mathrm{tot}}/R_{\mathrm{L}}$, where the load resistor $R_{\mathrm{L}}$ is constant. With a common supply voltage $V_{\mathrm{in}}$ supplied to all of the horizontal lines of the memristor crossbar array, the total resistance of the crossbar is given by $R_{\mathrm{tot}}= (\sum_{i,j}G_{ij}^{-1})^{-1}$, where $i,j$ represent the position of a memristor in the crossbar (see \cref{fig:setup}d). By programming $N_s$ distinct resistance states for each memristor, the feedback resistor $R_{\textrm{tot}}$ can take a maximum of $N_{s}^{N_m}$ distinct resistance values, where $N_m$ is the number of memristor in the crossbar circuit. Finally, we can define the theoretical voltage range $\Delta V$ and the optimal voltage resolution $\delta V$ of the memristor-based DC source:
\begin{equation}\label{eq:voltage_range}
    \Delta V=V_{max}-V_{min}=V_{\mathrm{in}}\frac{R_{\mathrm{HRS}}-R_{\mathrm{LRS}}}{R_LN_m}\\
\end{equation}
\begin{equation}\label{eq:voltage_res}
    \delta V = \Delta V/N_s^{N_m}
\end{equation}
where $R_{\mathrm{LRS}}$ and $R_{\mathrm{HRS}}$ are the resistance of the memristor in the lowest resistance state and the highest resistance state, respectively.\linebreak

\section{Thermal budget and power dissipation}\label{section_thermal}
The power dissipated by the circuit in the dilution fridge is an important consideration when designing cryogenic control electronics for quantum systems. For the memristor-based DC source, the two sources of dissipation are the operational amplifier and the memristors in the feedback loop placed at \SI{4.2}{\kelvin} (see \cref{fig:setup}c). On the one hand, the most recent custom cryogenic operational amplifier from Ref. \cite{cryo_ampli_le_guevel_2020} dissipates \SI{1}{\micro\watt}, while commercial amplifiers characterized at \SI{4.2}{\kelvin} are expected to dissipate $\approx$ \SI{50}{\micro\watt} \cite{Proctor2015}. \linebreak
On the other hand, the power dissipated by the memristors is introduced by the Joule heating proportional to the current flowing $V_{\textrm{in}}/R_L$ in the feedback loop. This dissipation is maximized when the memristor-based DC source delivers the maximum output voltage:
\begin{equation}
    P_{\textrm{mem}}^{\textrm{max}}=\left(\frac{V_{\textrm{in}}}{R_L}\right)^2R_{\textrm{tot}}^{\textrm{max}}
\end{equation}
where $R_{\textrm{tot}}^{\textrm{max}}=R_{\mathrm{HRS}}/N_m$ the maximum feedback loop resistance.
Our current memristors feature a \SI{16}{\kilo\ohm}-HRS, leading to a power dissipation of \SI{1.77}{\milli\watt} for a supply voltage $V_{\textrm{in}}=$ \SI{1}{\milli\volt} required to achieve a \SI{1}{\volt} supply range. Hence, the total power dissipation of a single memristor-based DC source is in the order of a milliwatt.  If we assume that this power is entirely dissipated on-chip, then an approximately 800 memristor-based DC source could be placed at the \SI{4.2}{\kelvin} stage assuming \SI{1.5}{\watt} cooling power.
This number is currently bottlenecked by the low HRS reachable with our memristor devices. Increasing the HRS would allow us to reduce the input voltage $V_{\textrm{in}}$, and thus the power dissipated, while keeping a \SI{1}{\volt}-range.\linebreak

\begin{figure*}[htbp] 
\center{\includegraphics[width=1\linewidth]{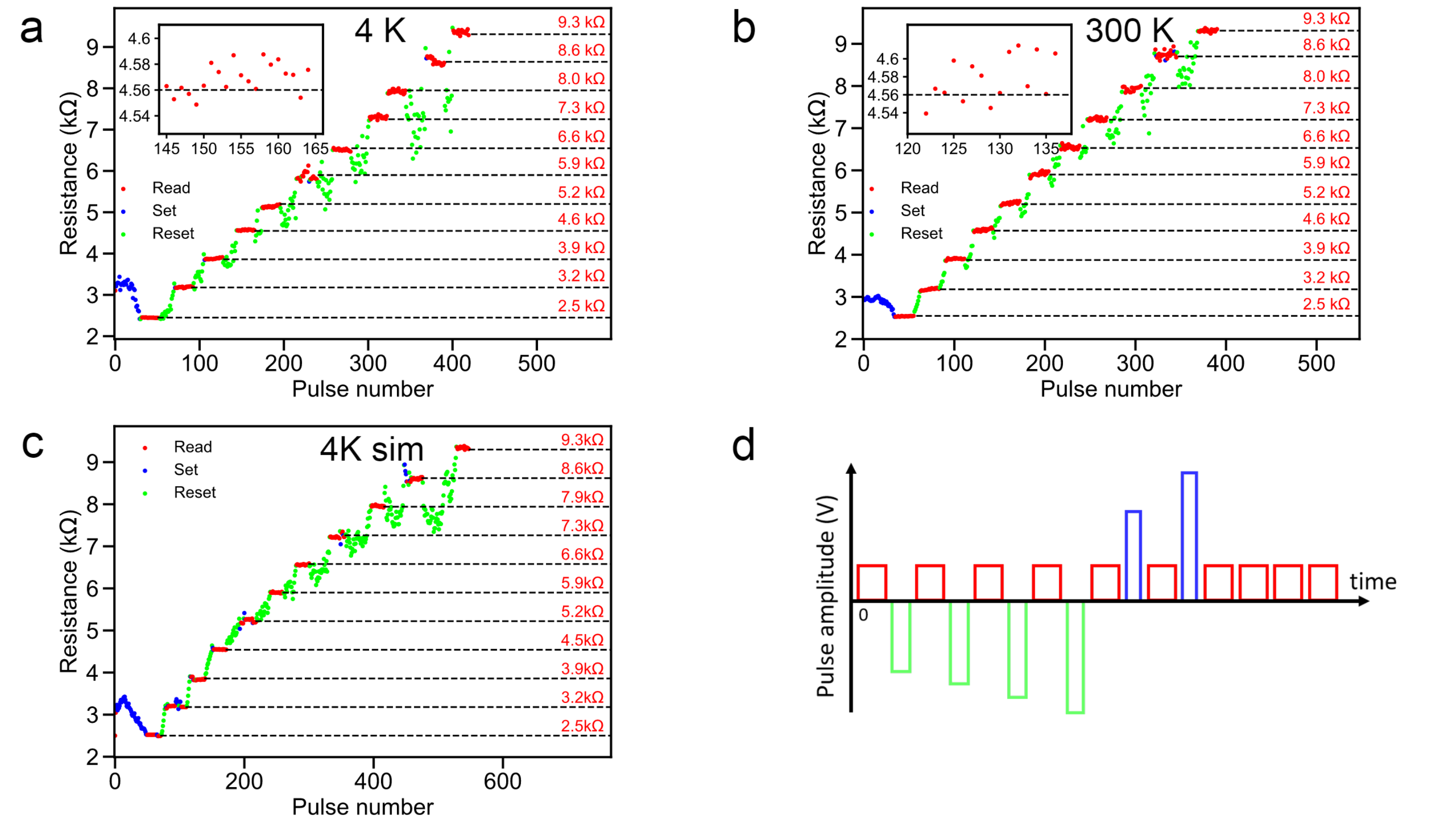}}
\caption{\textbf{Multilevel resistance state programming of a $\textrm{TiO}_{\textrm{2}}$-based memristor.} \textbf{a.} Programming 11 distinct resistance states within an accuracy of 1\% at \SI{4.2}{\kelvin}. Negative pulses are applied to the memristor electrodes to increase its resistance, while positive pulses decrease its resistance. These operations are, respectively, defined as RESET and SET. $V_p^{\mathrm{min}}=\SI{0.4}{\volt}$ and $V_n^{\mathrm{min}}=\SI{-0.6}{\volt}$ were chosen as initial pulse amplitudes for the multilevel programming, along with an amplitude step of \SI{0.02}{\volt}. The inset is a zoom-in of the \SI{4.56}{\kilo\ohm} resistance state using the same axis of the main plot. \textbf{b.}  Multilevel programming of 11 distinct resistance states within an accuracy of 1\% at room temperature. The initial pulse amplitudes that were chosen are $V_p^{\mathrm{min}}=\SI{0.7}{\volt}$ and $V_n^{\mathrm{min}}=\SI{-0.9}{\volt}$, with an amplitude step of \SI{0.02}{\volt}. The inset is a zoom-in of the \SI{4.56}{\kilo\ohm} resistance state showing sub-1\% reading variability. \textbf{c.} Simulation of a multilevel resistance state programming of a memristor placed at \SI{4.2}{\kelvin} based on the Data-Driven model \cite{Messaris_data_driven} fitted on our cryogenic experimental data. We used $V_p^{\mathrm{min}}=\SI{0.5}{\volt}$, $V_n^{\mathrm{min}}=\SI{-0.7}{\volt}$ and an amplitude step of \SI{0.02}{\volt} \textbf{d.} Typical pulse scheme used for the multilevel programming. The pulse width is fixed to \SI{200}{\nano\second} for write pulse (green or blue pulses) and to  \SI{10}{\micro\second} for reading pulse (red pulses). The writing amplitude varies linearly based on the previous measured resistance, while the reading amplitude is set to \SI{0.2}{\volt}.}
\label{fig:multilevel}
\end{figure*}

\begin{figure*}[htbp] 
\center{\includegraphics[width=1\linewidth]{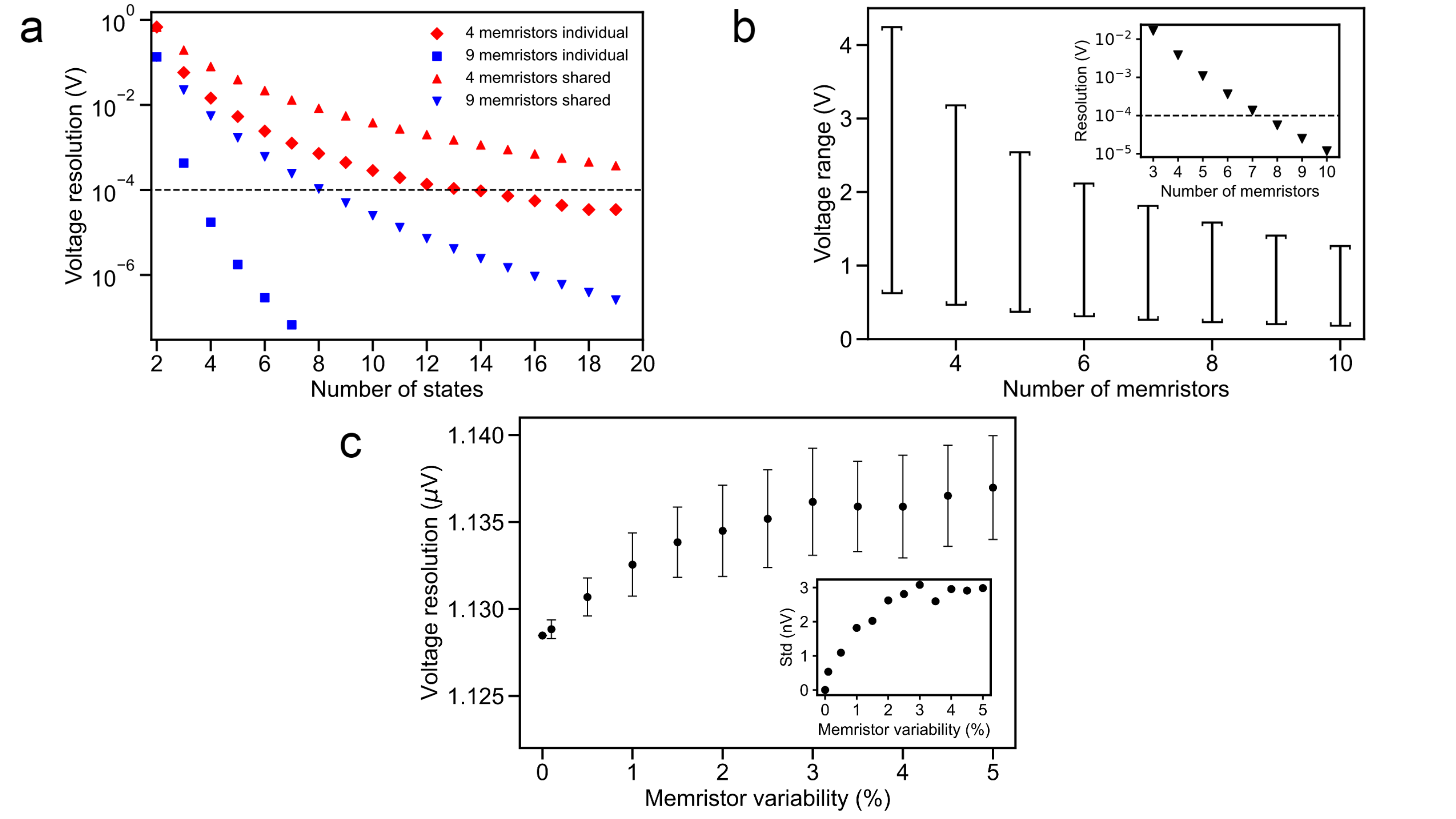}}
\caption{\textbf{Simulated performances of the memristor-based programmable DC source.} \textbf{a.} Evolution of the voltage resolution of the memristor-based DC source with respect to the number of states and the number of memristors in the circuit. The horizontal dashed-black line highlights the \SI{100}{\micro\volt} resolution target set to correctly reach an electronic regime in a quantum dot. A supply voltage $V_{\mathrm{in}}=\SI{1}{\milli\volt}$, a resistance load $R_L=\SI{1}{\ohm}$, a minimum resistance $R_{\mathrm{LRS}}=\SI{1.8}{\kilo\ohm}$ and a maximum resistance $R_{\mathrm{HRS}}=\SI{16}{\kilo\ohm}$, both inherited from the fitted cryogenic model, were used for these simulations. \textbf{b.} Maximum voltage range and optimal resolution for a given number of memristors. The number of distinct resistance states is set to five states for each memristor. As the number of memristors in the circuit increases, the voltage range shrinks while the voltage resolution improves leading to a resolution/range trade-off. \textbf{c.} Impact of the memristor variability on the DC source voltage resolution using a circuit of nine memristors with five states. The simulations were repeated 100 times for each variability value to allow for statistical analysis. The voltage resolution decreases logarithmically with the variability. The bottom inset shows the evolution of the standard deviation with respect to the memristor variability, which also follows a logarithmic dynamic. }
\label{fig:circuit_simulation}
\end{figure*}

\begin{figure*}[!t] 
\center{\includegraphics[width=0.7\linewidth]{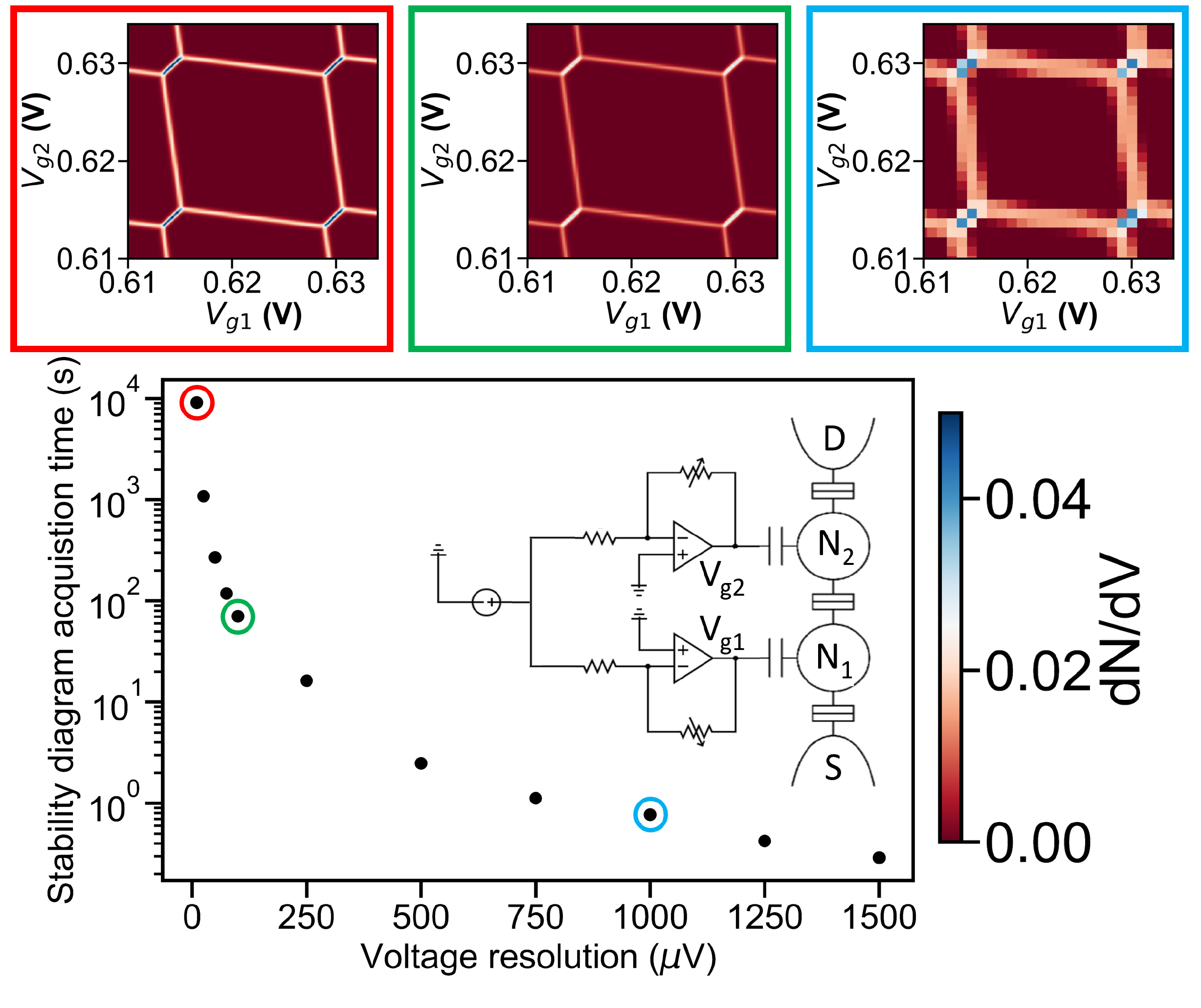}}
\caption{\textbf{Simulation of the co-integration of a double quantum dot and the memristor-based DC source.} Stability diagram acquisition time required with respect to the voltage resolution using two memristor-based cryogenic DC sources. The three color-framed insets show, respectively, the stability diagram obtained for a \SI{10}{\micro\volt}, \SI{100}{\micro\volt} and \SI{1}{\milli\volt} voltage resolution.  The co-integration simulations were conducted with a double quantum dot placed at \SI{10}{\milli\kelvin} to take into account thermal noise and the DC source at \SI{4}{\kelvin} using the cryogenic Data-Driven memristor model depicted in the inset. The parameters used for this set of simulations are summarized in \cref{tab:param_quantum} for the double quantum dot and \cref{tab:param_memristor} for the memristor in the \hyperref[Methods]{Supplementary materials}. }
\label{fig:quantum_simu}
\end{figure*}

\section{Experimental results}

Demonstrating reversible, non-volatile, and linear resistance programming of resistive memory devices at cryogenic temperatures would offer opportunities for hybrid memristor–CMOS cryogenic electronics. To accurately tune resistance states, we use a simple feedback algorithm from Ref. \cite{Alibart2012}. At each step, we apply a \SI{200}{\nano\second} write pulse of negative amplitude to increase the resistance of the device or a positive amplitude to decrease its resistance, followed by a \SI{1}{\micro\second}/\SI{200}{\milli\volt} read pulse to evaluate the current resistance state of the memristor (see \cref{fig:multilevel}d). The write pulse amplitude is increased linearly by \SI{20}{\milli\volt}-step, starting from an initial positive amplitude $V_p^{\mathrm{min}}$ or negative amplitude $V_n^{\mathrm{min}}$ depending on the initial resistance state of the memristor. 
Once the desired resistance state is reached within a chosen tolerance, the resistance is read 20 times to ensure state stability (see insets of \cref{fig:multilevel}a, b).\linebreak

Measurements were conducted on the same sample at \SI{4.2}{\kelvin} and room temperature using a Keysight B1500A semiconductor analyzer with a \SI{200}{\mega samples/\second} waveform generator module (WGFMU) coupled with a Lakeshore CRX-VF cryogenic probe station (see \cref{fig:multilevel}a and b). 
The TiO$_{\textrm{2}}$-based memristors were formed beforehand at \SI{300}{\kelvin}. We demonstrate 11 distinct resistance states at \SI{4.2}{\kelvin} and \SI{300}{\kelvin}, with approximately 50 write pulses for each state with a programming accuracy of 1\%. We observe that the resistance can decrease instead of increasing when applying negative write pulses during the multilevel programming. This phenomenon is more important at larger resistance values and can be defined as write variability. Cryogenic multilevel programming of the memristor (See \cref{fig:multilevel}b) shows a larger resistance decrease, which could be imputed to the larger temperature gradient in the vicinity of the conductive filament \cite{Strukov2012}. Moreover, at cryogenic temperatures, the competition between thermal- and field-activated effects is increased and could lead to additional variability during the thermally-activated RESET operation (increasing of the memristor resistance) \cite{Beilliard2020IOP, Govoreanu_cryo_pulse}. One can also note reading variability from the inset of \cref{fig:multilevel}a and b, which can be associated with thermal noise induced by the memristor conduction modes. 
We evaluate a sub-1\% read variability on our devices, which allows for precise and stable resistance states (see \hyperref[fig:varia]{Supplementary Fig. 3}  for the characterization of the reading variability).  A better programming accuracy can be reached using a smaller voltage step for pulse amplitude at the cost of programming time (i.e., more write pulses). Finally, we simulated the multilevel programming of our TiO$_2$-based memristors (See \cref{fig:multilevel}c) by fitting a Data-Driven model \cite{Messaris_data_driven} replicating the pulsed response transient of a resistive memory (see \hyperref[fig:datadriven]{Supplementary Fig. 2}) for the fitting) that will enable the design of cryogenic memristor circuits. This model closely replicates the programming of the resistance states, which allows high-fidelity hardware-based simulations of the DC source in the following section.\par

\section{Simulations and discussions}

Using the Data-Driven memristor model \cite{Messaris_data_driven} parameters fitted on our experimental data at \SI{4.2}{\kelvin}, we will next benchmark the performance of the memristor-based DC source by varying the number of memristors in the circuit and the number of distinct resistance states. We initially consider the circuit with no read or write variability to assess the optimal voltage resolution and range achievable. 

\Cref{fig:circuit_simulation}a shows the evolution of the voltage resolution of the memristor-based DC source with respect to the number of states and the number of memristors. 
A higher number of resistance states leads to a better voltage resolution, which is limited by the maximum number of distinct resistance states achievable with the memristor used.
By programming five states on nine memristors, we ideally reach a voltage resolution of $\approx$ \SI{1}{\micro\volt}, which is well below the \SI{100}{\micro\volt} target resolution required for QDs biasing. This five-states distribution is therefore used for the following simulations. 
\Cref{fig:circuit_simulation}b specifies the basic electrical characteristics of the DC-source with respect to the number of memristors used. This figure shows the trade-off between the voltage resolution and the range as the number of memristors increases, this trend is explained by equations \eqref{eq:voltage_range} and \eqref{eq:voltage_res}. For an increasing number of memristors, the voltage range decreases while the resolution increases exponentially. To comply with the \SI{100}{\micro\volt} voltage resolution and \SI{1}{\volt}-range requirements for quantum dot biasing, a circuit of at least eight memristors is needed if the number of states is fixed to five, which eases the multilevel programming operation. In the remainder of this paper, we consider a nine-memristor circuit to utilize the denser and more common 3$\times$3 memristor-crossbar footprint.\linebreak

To properly evaluate the memristor-based DC electrical specifications, we investigated the impact of memristor non-idealities \cite{Adam2018} on the DC source output voltage. \Cref{fig:circuit_simulation}c shows the voltage resolution of a nine-memristor five-state circuit with respect to the memristor variability. We chose the memristor variability as the combination of the read and write variabilities observed with our devices and implemented it as a percentage error of the resistance for each memristor. As this simulated memristor variability increases, the voltage resolution of the DC source remains almost constant. This suggests that the memristor's variability has a negligible impact on the voltage resolution of the memristor-based DC source. We also note a decrease in the reproducibility of the constant step size between two consecutive output voltages (see inset of \cref{fig:circuit_simulation}c). We can compare the averaged error induced by the memristor variability to the voltage noise of DC sources that are used in quantum dot setups,  typically in the order of 1 to 5\% of the target voltage (2.0\% for the Tektronix 5014C). \linebreak

The proposed DC source was programmed by reaching target resistance states for each memristor, leading to an associated output voltage. In the preceding simulations, we chose to consider the memristors as discrete multi-state resistors. This allowed for efficient programming because a simple feedback resistance algorithm can be used, as in \cite{Alibart2012}, to tune the whole crossbar array. Using this discrete programming approach, the output voltages depend on the resistance targets that we initially set, thus limiting our control of the output voltage. Instead, the memristor-based DC source can be controlled  to reach an arbitrary target voltage to enable a more precise control of the PGA output voltage by exploiting the fully analog behaviour of the memristors. We propose a servo algorithm, which will program the memristor crossbar to reach the target voltage accordingly. The crossbar resistance configuration is chosen in a pre-computed dataset, where the keys are target output voltages. We converge towards this optimal resistance configuration by performing a first coarse round of programming for each memristor to roughly reach the attributed resistance using the feedback algorithm  for multilevel programming (see \cref{fig:multilevel}a, b and c). During a second round of programming, the resistance of the last memristor of the crossbar is finely tuned with a smaller voltage step and programming tolerance to a new target resistance that is specifically computed to balance the programming inaccuracy of the initial fast programming made on the other memristors. This fine tuning method allows for memristor variability robustness and a constant \SI{100}{\micro\volt} step size within a 2.5\% voltage error using a nine-memristor circuit.  Controlling the memristor-based DC source to reach an arbitrary voltage offers a smaller voltage resolution when compared to the resistance-driven operation mode. However, the proposed memristor-based programmable DC source still satisfies the \SI{100}{\micro\volt} for the same circuit parameters---that is, nine memristors and a supply voltage $V_{\textrm{in}}=$\SI{1}{\milli\volt}---and is more suitable for \textit{in-situ} use of the DC source.
\linebreak

Now that we have explained the working principle of our memristor-based DC source and demonstrated how it can be operated to reach a target voltage, we will next simulate its co-integration  with a double quantum dot to validate its compatibility with quantum systems. Stability diagrams are used as a standard benchmark measurement to verify the basic performance of a quantum dot system. We simulated the electron transport through a double quantum dot (DQD) with respect to the voltages applied to its gates while biased by two memristor-based cryogenic DC source using equations from \cite{VDW_QD_transport} (see \hyperref[Methods]{Methods}). Two important metrics to consider for a QD biasing solution are the voltage resolution and the stability diagram acquisition time, which is estimated by counting the number of write and read pulses needed to program the memristor circuit during the simulation of the charge stability diagram. This acquisition time increases with better voltage resolution, leading to a trade-off between speed and accuracy (see \cref{fig:quantum_simu}). The insets of \Cref{fig:quantum_simu} show the derivative of a charge stability diagram to highlight the transition between the electronic regimes of the simulated DQD. We can identify the transition and occupations of the DQD for the three voltage resolutions---that is, \SI{1}{\milli\volt}, \SI{100}{\micro\volt} and \SI{10}{\micro\volt}---, thus demonstrating the control of a simulated DQD by the proposed memristor-based DC source.  Although the DC source can conveniently deliver voltage resolution from \SI{10}{\milli\volt} to below \SI{1}{\micro\volt}, a \SI{100}{\micro\volt} resolution is sufficient to clearly observe the honeycomb pattern and its triple-point. It also allows for a fast  \SI{70}{\second}-scanning  of the DQD stability diagram. Moreover, when scaling up the DC-sources for a large number of double quantum dots, the stability diagram measurement can be performed in parallel. This allows us to scan all double quantum dots within \SI{70}{} seconds using a \SI{100}{\micro\volt} resolution.
\linebreak

\section{Conclusion}

In conclusion, we propose a memristor-based cryogenic programmable DC source for scalable in-situ quantum-dot control. We demonstrate that this control approach fulfills the baseline requirements for quantum dot biasing (i.e., a \SI{100}{\micro\volt}-resolution over a \SI{1}{\volt} range) while keeping a limited number of memristors, allowing for a scalable biasing solution. We demonstrated the effective biasing of a simulated double quantum dot using a non-ideal TiO$_2$/Al$_2$O$_3$ memristor model, pointing out that memristor variability will not cause electronic regime changes that would be detrimental to quantum computation. We reported successful multilevel pulsed programming of TiO$_2$-based memristor at \SI{4.2}{\kelvin}. This demonstration is a first step towards more advanced cryogenic applications for resistive memories, such as cryogenic control electronics for quantum computers using this proposed in-situ memristor-based DC source.
Moreover, the scalability of this solution is limited by the total power dissipation of the DC source but still hints at a feasible integration of a few hundred of DC sources at \SI{4.2}{\kelvin}, where the cooling power is on the order of a watt. More resistive memristors would allow us to decrease the power dissipation by limiting the current flowing in the feedback loop.  Further works remain to be done regarding the thermal load introduced by memristors while they are being programmed to ensure cryo-compatibility. However, this additional heat load is limited to the bias tuning phase because memristors are non-volatile and the supply is maintained once programmed.\linebreak
Furthermore, the recent demonstrations of silicon spin qubits operated at temperatures above \SI{1}{\kelvin} \cite{Yang2020,Petit2020} paves the way towards on-chip co-integration of 'hot' quantum dots and memristor-based control electronics. This approach would allow us to overcome the wiring bottleneck that the community will face when scaling-up quantum computers. \\
\linebreak

\textbf{Data availability}

The code to control the Keysight B1500 that was used to perform the measurements is available at \href{https://github.com/3it-nano/B1530A_control}{GitHub} \cite{MounyB15002021} and the Python library that was developed to simulate the control performed by the proposed DC source is available at \href{https://github.com/3it-nano/QDMS}{GitHub} \cite{MounyQDMS2021}. The figures were created using Matplotlib \cite{Hunter_plt}
Data supporting this work will be uploaded to an online repository.\\
\linebreak
\textbf{Acknowledgements}
This work was supported by Natural Sciences and Engineering Research Council of Canada (NSERC). This research was undertaken thanks in part to funding from the Canada First Research Excellence Fund. We acknowledge financial supports from the EU: ERC-2017-COG project IONOS (\# GA 773228). We would like to acknowledge Christian Lupien, Edouard Pinsolle and the Institut Quantique for their assistance with the electrical characterisation at cryogenic temperatures. LN2 is French-Canadian joint International Research Laboratory (IRL-3463) funded and co-operated by CNRS, Université de Sherbrooke, Université de Grenoble Alpes (UGA), École Centrale Lyon (ECL) and INSA Lyon. It is supported by the Fonds de Recherche du Québec Nature et Technologie (FRQNT). We would also like to thank the IEMN cleanroom engineers for their support with the device fabrication.
\\

\noindent\textbf{Author contributions}
P.A.M. and S.G. performed the experiment. A.E.M., R.D., P.G. and S.E. fabricated the memristor devices. P.A.M. analysed and post-processed the measurement and simulation data. P.A.M., S.G. and M.A.R. developed the simulation framework Y.B., D.D. and M.P.L. conceived and supervised the project. P.A.M. wrote the manuscript with input from all of the authors.\\
\linebreak
\textbf{Competing interests}
The authors declare no competing interests.
\bibliographystyle{unsrt}
\bibliography{references}
\pagebreak
\clearpage
\section*{Methods}\label{Methods}

\setlength{\parindent}{0em}
\textbf{Fabrication of a TiO${_\textrm{2}}$-based memristor.} The sample that was used for the measurements presented in this article followed our fabrication process of CMOS-compatible Al$_{\mathrm{2}}$O$_{\mathrm{3}}$/TiO$_{\mathrm{2-x}}$ memristor devices described in \cite{mesoudy2021CMOS}, with two main differences: the TiO$_{\mathrm{2}}$ layer is \SI{15}{\nano\metre} thick, and the electrodes are \SI{200}{\nano\metre} wide.
\\

\textbf{Electrical characterizations} All of the electrical characterizations were conducted on a Keysight B1500 using the WGFMUs module enabling the generation of the \SI{200}{\nano\second} write pulses paired with a Lakeshore CRX-VF probe station allowing for measurements down to approximately \SI{2}{\kelvin}. The WGFMUs are controlled using a GPIB-USB connection along with a custom C\texttt{++} script based on a Keysigh library \cite{MounyB15002021}. The script allows us to perform multilevel resistance programming, as well as the measurement required for the fitting of our simulation model.  
\\

\textbf{Read variability assessments.} The read variability of a memristor is measured by  biasing the device at the voltage that is typically used for read pulse (i.e., \SI{0.2}{\volt}). The current flowing the through the memristor is measured every \SI{10}{\micro\second} (sampling rate $f_s = \SI{100}{\kilo\hertz}$) for \SI{10}{\second} because the B1500 analyzer is limited to a maximum of four MSamples. Each sample point is the result of an averaging over \SI{10}{\micro\second}. This approach allows us to smoothen the read variability leading to a smaller standard deviation in the Gaussian fit, while replicating the action of a read pulse used in multilevel programming measurements. The experimental data can be fitted by a Gaussian to estimate the reading variability of the device. We conducted these measurements on the same memristor at room temperature and \SI{4}{\kelvin} (see \hyperref[fig:varia]{Supplementary Fig. 3}). The coefficients extracted from the linear regression applied to the post-processed measurement data are used to complete our simulation model and implement variability to the emulated memristor.
\\

\textbf{Simulations.} 
The simulation framework was developed as a two parts (classical electronics and quantum) simulator. The classical simulations are based on Kirchhoff's laws, the differential amplifier equation and the data driven model giving the memristor resistance with respect to the write pulse applied \cite{Messaris_data_driven}. We added additional parameters to the data driven model to replicate basic non-idealities of the memristor, such as read and write variability, to more accurately predict the DC source behaviour. The quantum simulations are limited to the plotting of stability diagrams. We used the equations of the electron transport through a double quantum dot in reference \cite{VDW_QD_transport} to determine the double quantum dot occupation with respect to the gate voltages ($V_{\textrm{g1}}$ and $V_{\textrm{g2}}$) under the assumption of solely Coulomb blockade. The stability diagrams presented in this article are plotted by derivating the dot occupation with respect to both variables ($V_{\textrm{g1}}$ and $V_{\textrm{g2}}$). The quantum simulation can be ran for different existing quantum dots. We used capacitance parameters from the University of New South Wales from reference \cite{Hwang_2018} as a standard but implemented quantum dot parameters from other leading groups \cite{Xue2021, CEA_LETI_QD, Sandia_QD, UCL_QD, Zajac_2018}. These parameters were extracted by post-processing available stability diagrams using electron transport equations. The quantum simulation is temperature-dependent, which allows for the simulation of hot silicon quantum dots \cite{Petit2020, Yang2020}. The memristor model can also be changed according to the working temperature, either at \SI{4}{\kelvin} or at room temperature (all of the simulation results that are presented in this article used the \SI{4}{\kelvin} memristor model).
\\



\renewcommand{\figurename}{Supplementary Fig.}
\setcounter{figure}{0}
\setcounter{table}{0}
\begin{table}[h]
\centering
\begin{tabular}{p{2.8cm}p{3.2cm}p{2.0cm}}
 \hline
 Parameter& Value& Unit\\
 \hline
 $\textrm{C}_{\textrm{g1}}$ & $1.03\times 10^{-18}$& F\\ 
 $\textrm{C}_{\textrm{g2}}$ & $1.03\times 10^{-18}$& F\\
 $\textrm{C}_{\textrm{m}}$ & $4.0\times 10^{-19}$& F\\
 $\textrm{C}_{\textrm{L}}$ & $5.0\times 10^{-18}$& F\\
 $\textrm{C}_{\textrm{R}}$ & $5.0\times 10^{-18}$& F\\
  T& $10^{-3}$& K\\
 \hline
\end{tabular}
\caption{\textbf{Parameters of the double quantum dot for the simulation.} The quantum dot parameters were extracted by post-processing stability diagrams from reference \cite{Hwang_2018}.}
\label{tab:param_quantum}
\end{table}
\begin{table}[b]
\centering
\begin{tabular}{p{4.0cm}p{4.0cm}}
 \hline
 Parameter& Value\\
 \hline
 $\textrm{R}_{\textrm{on}}$ & \SI{1.8}{\kilo\ohm} \\ 
 $\textrm{R}_{\textrm{off}}$ & \SI{16.0}{\kilo\ohm}\\
 $\textrm{A}_{\textrm{p}}$ & $2.57\times 10^{2}$\\
 $\textrm{A}_{\textrm{n}}$ & $-1.01\times 10^{2}$\\
 $\textrm{t}_{\textrm{p}}$ & $-1.81$\\
 $\textrm{t}_{\textrm{n}}$ & $-5.53\times 10^{-1}$\\
$\textrm{a}_{\textrm{p}}$ & $3.33\times 10^{-1}$\\
$\textrm{b}_{\textrm{p}}$ & $1.87$\\
$\textrm{a}_{\textrm{n}}$ & $3.33\times 10^{-1}$\\
$\textrm{b}_{\textrm{n}}$ & $1.87$\\
$\textrm{r}_{\textrm{p0}}$ & $9.23\times 10^{3}$\\
$\textrm{r}_{\textrm{p1}}$ & $-6.34\times 10^{3}$\\ 
$\textrm{r}_{\textrm{n0}}$ & $-4.59\times 10^{3}$\\
$\textrm{r}_{\textrm{n1}}$ & $-1.60\times 10^{4}$\\
 \hline
\end{tabular}
\caption{\textbf{Fitted parameters of a memristor.} These parameters were fitted on a \SI{4.2}{\kelvin} pulsed measurement performed on a memristor. The fitting of this Data-driven model is presented at the \hyperref[fig:datadriven]{Supplementary Figure 2} and extensively presented in Reference \cite{Messaris_data_driven}}
\label{tab:param_memristor}
\end{table}
\pagebreak

\begin{figure*}[t]  
    \center{\includegraphics[width=1\linewidth]{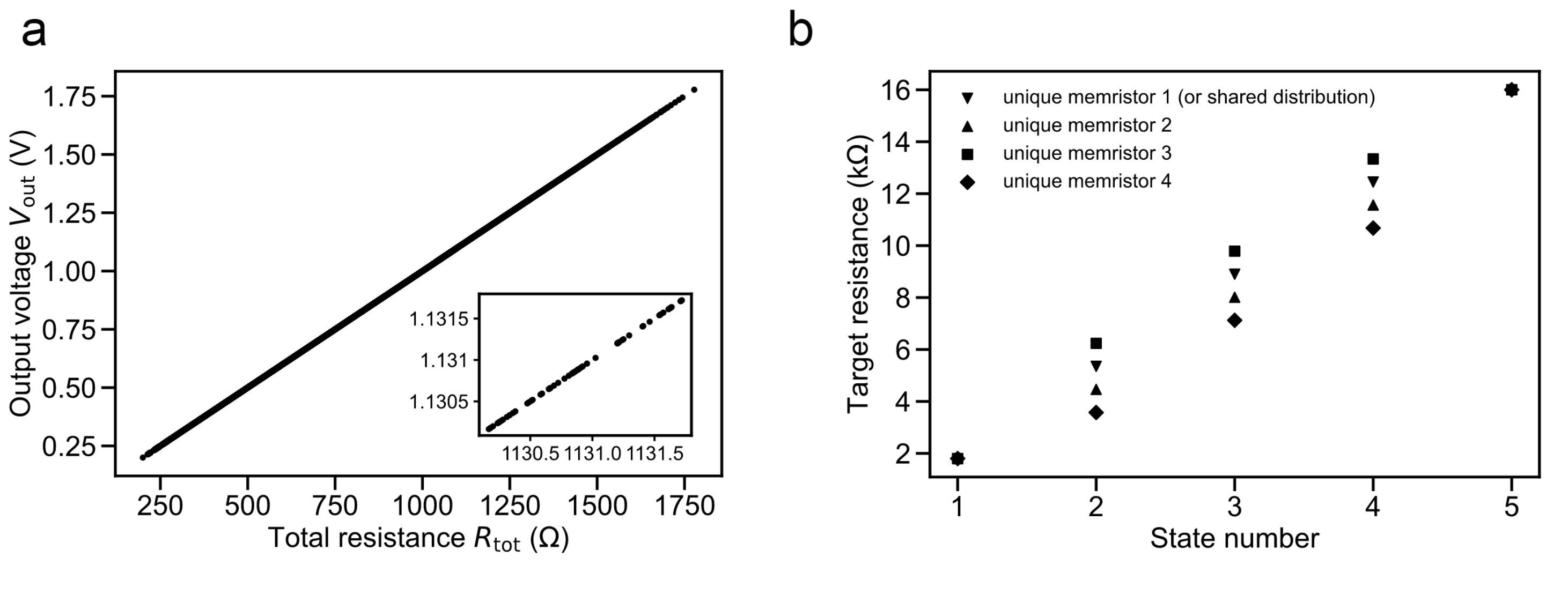}}
    \caption{\textbf{Voltage output of the memristor-based  DC source.} \textbf{a.}\, Detailed voltage output of a 9-memristor  DC source with a five-state unique distribution for each memristor. The voltage step between consecutive voltages is not constant, as the zoom-in in the bottom inset emphasizes.  \textbf{b.}\, Schematic representation of a five-state unique distribution for a four-memristor DC source. The minimum and maximum target resistance are equal for all of the distributions because these values are tied to the memristor's properties. The distribution of memristor 1 (down triangle) represents the default distribution when the memristors share the same distribution.}
    \label{fig:schematic}
\end{figure*}

\begin{figure*}[t] 
    \center{\includegraphics[width=1\linewidth]{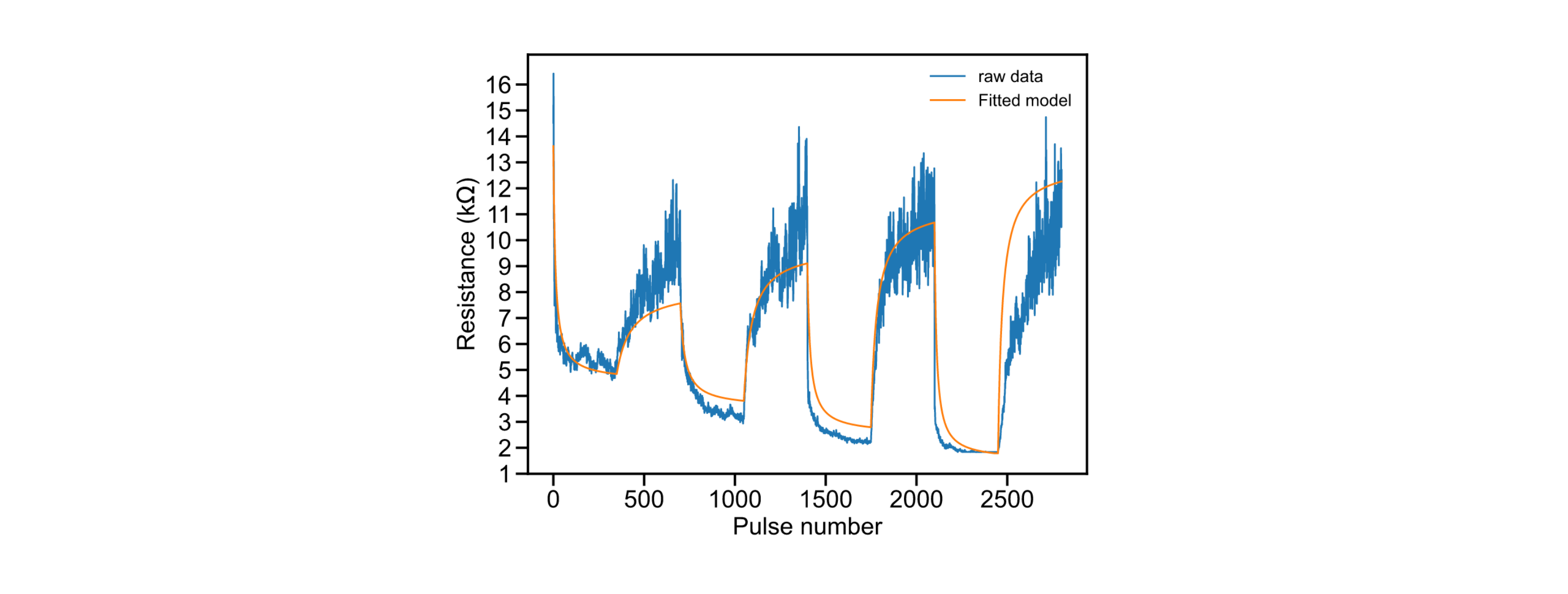}}
    \caption{\textbf{Fitting of the data driven memristor model \cite{Messaris_data_driven} on measurements performed at 4.2 K}. We apply multiple pulse trains of N write/read pulse sequences to the memristor. The pulse amplitude is increased linearly after each pulse train. During a pulse train, the polarity of the pulse is changed after N/2 write/read sequences. The fitting allows us to extract the parameters that were used to model the memristor pulsed behaviors used in our simulations. We used a write pulse width $t_{w}=\SI{200}{\nano\second}$ and a write pulse amplitude $A_w\in$ [0.8, -0.8, 0.9, -0.9, 1.0, -1.0, -1.1]. 350 read/write pulse sequences were applied for each amplitude. We used the same read pulses parameters as the multilevel programming: $t_r = \SI{10}{\micro\second}$ and $A_r=\SI{0.2}{\volt}$. The fitted parameters  are used in the simulations presented in this article and are available on our GitHub repository \cite{MounyQDMS2021} and at \cref{tab:param_memristor}. 
    }
    \label{fig:datadriven}
\end{figure*}


\begin{figure*}[t]  
    \center{\includegraphics[width=1\linewidth]{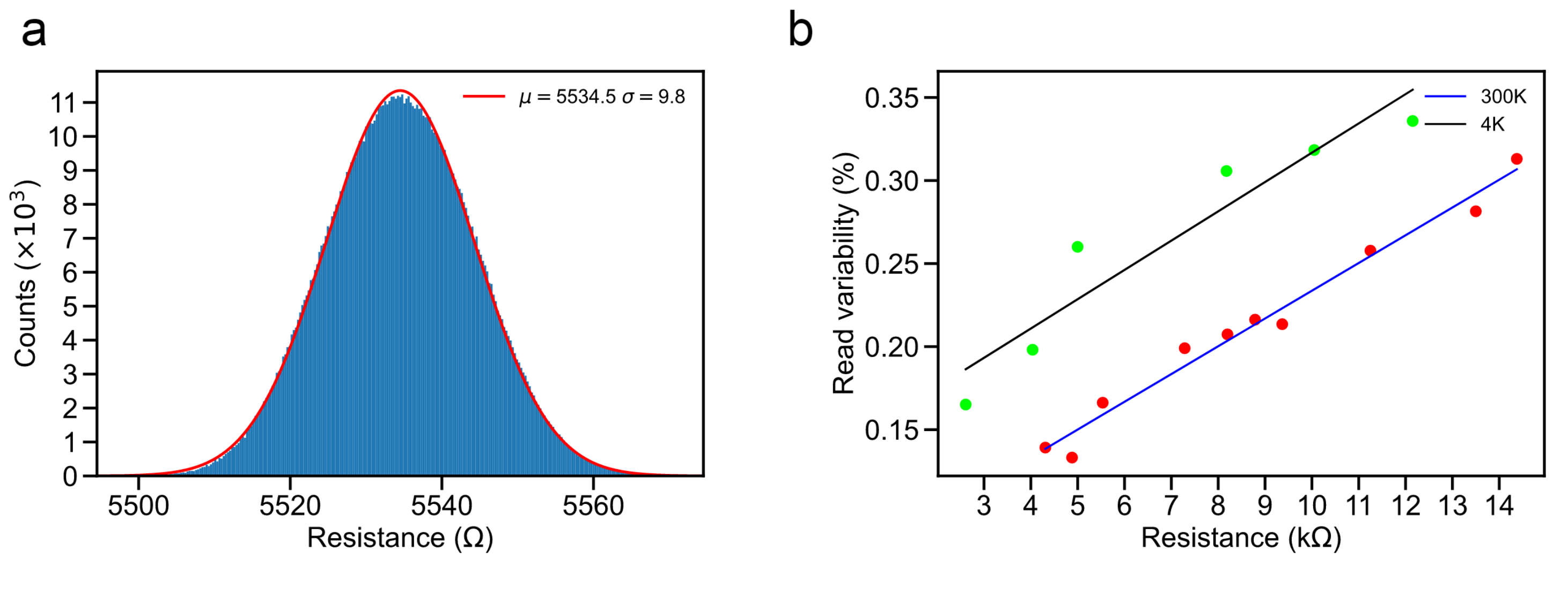}}
    \caption{\textbf{Reading variability of a memristor crosspoint} \textbf{a.} Gaussian fit of a read variability measurement at \SI{300}{\kelvin}. From this fit, we can extract the read variability of the memristor for a given resistance state. \textbf{b.} We can fit the variability by reproducing the previous measurement and fit for various resistance states of a single memristor, which increases linearly with the resistance between the LRS and HRS resistance states. One can notice a slight offset in read variability between cryogenic and room temperatures, suggesting that cooling down memristors will not decrease the read variability.
    }
    \label{fig:varia}
\end{figure*} 
\pagebreak

\end{document}